\documentclass[universe,article,accept,moreauthors,pdftex]{mdpi}

\newcommand{\ba}{\begin{eqnarray}}
\newcommand{\ea}{\end{eqnarray}}
\newcommand{\be}{\begin{equation}}
\newcommand{\ee}{\end{equation}}

\usepackage{amssymb}
\usepackage{soul}
\usepackage{color}

\def\be{\begin{equation}}
\def\ee{\end{equation}}
\def\bea{\begin{eqnarray}}
\def\eea{\end{eqnarray}}

\firstpage{1}
\pubvolume{7}
\issuenum{7}
\articlenumber{227}
\pubyear{2021}
\copyrightyear{2021}
\externaleditor{Academic Editors: Sebastian Bahamonde and Jackson Levi Said} 
\datereceived{28 May 2021}
\dateaccepted{2 July 2021}
\datepublished{5 July 2021}
\hreflink{https://doi.org/10.3390/\linebreak universe7070227} 

\Title{Gravitationally Induced Particle Production through a Nonminimal Torsion--Matter Coupling} 

\TitleCitation{Gravitationally Induced Particle Production through a Nonminimal Torsion--Matter Coupling}

\Author{Tiberiu Harko 
$^{1,2,3}$, \mbox{Francisco S. N. Lobo $^{4,5,}$*\orcidA{}} and Emmanuel N. Saridakis 
$^{6,7,8}$\orcidB{}}

\AuthorNames{Tiberiu Harko, Francisco S. N. Lobo and  Emmanuel N. Saridakis}

\AuthorCitation{Harko, T.; Lobo, F.S.N.; Saridakis, E.N.}

\address{%
$^{1}$ \quad Astronomical Observatory, 19 Ciresilor Street, 400487 Cluj-Napoca, Romania; tiberiu.harko@aira.astro.ro\\

$^{2}$ \quad Department of Physics, Babes-Bolyai University, Kogalniceanu Street,
400084 Cluj-Napoca, Romania\\
$^{3}$ \quad School of Physics, Sun Yat-Sen University, Xingang Road, Guangzhou 510275, China\\
$^{4}$ \quad Instituto de Astrof\'{i}sica e Ci\^{e}ncias do Espa\c{c}o, Faculdade de Ci\^encias da Universidade de Lisboa, Edif\'{i}cio C8, Campo Grande, PT1749-016 Lisbon, Portugal\\
$^{5}$ \quad Departamento de F\'{i}sica, Faculdade de Ci\^{e}ncias, Universidade de Lisboa, Edifício C8, Campo Grande, PT1749-016 Lisbon, Portugal\\
$^{6}$ \quad National Observatory of Athens, Lofos Nymfon, 11852 Athens, Greece; msaridak@noa.gr\\
$^{7}$ \quad CAS Key Laboratory for Researches in Galaxies and Cosmology,
Department of Astronomy, University of Science and Technology of China,
Hefei 230026, China\\
$^{8}$ \quad School of Astronomy, School of Physical Sciences,
University of Science and Technology of China,\linebreak Hefei 230026, China

}

\corres{Correspondence: fslobo@fc.ul.pt; Tel.: +351-217-500-484; Fax: +351-217-500-977}

\abstract{We investigate the possibility of gravitationally generated particle production via the mechanism of  nonminimal torsion--matter coupling. An intriguing feature of this theory is that the divergence of the matter energy--momentum tensor does not vanish identically. We explore the physical and cosmological implications of the nonconservation of the energy--momentum tensor by using the formalism of  irreversible thermodynamics of open systems in the presence of matter creation/annihilation.   The particle creation rates,  pressure, and the expression of the comoving entropy are obtained in a covariant  formulation and discussed in detail.  Applied together with the gravitational field equations, the thermodynamics of open systems lead to a generalization of the standard $\Lambda$CDM cosmological paradigm, in which the particle creation rates and  pressures  are effectively considered as components of the cosmological fluid energy--momentum tensor. We consider specific models, and we show that  cosmology with a torsion--matter coupling can  almost perfectly reproduce the $\Lambda$CDM scenario, while it additionally gives rise to particle creation  rates, creation pressures, and entropy generation through gravitational matter production in both low and high redshift limits.}

\keyword{particle production; irreversible thermodynamics; nonminimal torsion--matter coupling; modified theories of gravity}

\begin{document}


\section{Introduction}

General Relativity has been established as the theory of gravitational
interactions for over a century, being consistent with all experiments and
being able to describe a huge set of observational results. Nevertheless,
there are two justified motivations that lead to  the construction of
its modifications and extensions. The
first motivation is based on  cosmological grounds and aims to offer a
successful description of the two phases of accelerated expansions, without
facing the cosmological constant problem or without the need to introduce extra
fields/fluids, such as the inflaton and/or the dark energy \mbox{sectors
\cite{Nojiri:2006ri,Clifton:2011jh,Capozziello:2011et}.} Moreover, one hopes to
alleviate the two possible
tensions that recently puzzle  $\Lambda$CDM cosmology, namely the $H_0$
\cite{DiValentino:2020zio}  and the $\sigma_8$ \cite{DiValentino:2020vvd}
tensions.  The second  motivation arises from theoretical considerations and
considers general relativity as the low-energy limit of a richer and more
fundamental theory that would be renormalizable and closer to the full quantum
gravitational one \cite{Stelle:1976gc,Gross:1986mw,Biswas:2011ar}.

In order to construct modified gravity theories, one starts from the
Einstein--Hilbert Lagrangian and extends it in all possible ways \cite{Saridakis:2021lqd}.  However, one
can equally, as well, start from the equivalent gravitational description in terms
of torsion, namely from the teleparallel equivalent of general relativity \cite{TEGR,TEGR2,Hayashi:1979qx,Maluf:2013gaa}, and
extend the corresponding Lagrangian, given by the torsion scalar $T$. In this way, we can
obtain $f(T)$ gravity \cite{Bengochea:2008gz,Cai:2015emx},   $f(T,T_G)$ \mbox{gravity
\cite{Kofinas:2014owa},} scalar-torsion theories \cite{Geng:2011aj}, $f(T,B)$
gravity \cite{Bahamonde:2015zma}, etc.
Along these lines, one can further extend the torsional
formulation in theories with nontrivial couplings between gravity and
the matter sector, such as in $f(T,\overset{\text{em}}{T})$ (where
$\overset{\text{em}}{T}$ is the trace of the matter energy--momentum tensor) \cite{Harko:2014aja}, or in
theories with a nonminimal coupling between the torsion scalar and the matter
Lagrangian \cite{Harko:2014sja}. These theories prove to have interesting
cosmological \mbox{applications
\cite{Harko:2014sja,Jawad:2016zwj,Bahamonde:2017ifa,DAgostino:2018ngy,
Gonzalez-Espinoza:2018gyl}.} One important feature of these classes of
theory is that the matter and effective dark energy sectors are not separately
conserved, since there exists an effective interaction between torsional geometry
and matter.

On the other hand, irreversible thermodynamics and thermodynamics of open
systems is a widely studied field since it is useful in various applications
\cite{Pri0,Pri,Cal,Lima,Ha1, Ha2,Mim,Pav,Bak}. One of them is related to the possibility of
particle creation and  matter production on   cosmological scales, due to the
appearance of a ``heat''-type term  which corresponds to the internal energy of the system. This   matter
production  can be equivalently
described through the addition of
an effective bulk viscous pressure into the energy--momentum tensor of the
cosmological
fluid \cite{Zeld,Mur,Hu}.

In the present work, we are interested in investigating the possibility of gravitationally induced particle production
through the mechanism of a nonminimal torsion--matter coupling. In particular,
since, in such theories, we naturally obtain the nonconservation of the ordinary
matter energy--momentum tensor, we can interpret it as   an irreversible matter
creation process, describing the generation of particles due to   torsion.
Additionally, particle production   acts as an
entropy source, too, which generates an effective entropy flux. Hence, in
the presence of the torsion--matter coupling,  the temperature evolution of the
thermodynamic systems is also modified.

The structure of this paper is as follows. In Section \ref{matter}, we present the
theory of a nonminimal torsion--matter coupling, and we apply it in a
cosmological framework, extracting the relevant equations. In Section
\ref{sect3}, we briefly review the thermodynamics of open systems, focusing on
the second law of thermodynamics, the particle generation rates,  and the
creation pressure, in addition to relating the
particle creation and bulk viscosity. Then, in Section \ref{sect4},
we investigate the  cosmological evolution and  particle generation in the
case of  nonminimal torsion--matter coupling gravity. Finally, in Section
\ref{sect5} we summarize the obtained results and conclude.

\section{ \boldmath{$f(T)$} Gravity with Nonminimal Torsion--Matter Coupling}\label{matter}

In this section, we briefly review  the $f(T)$ gravity formalism with nonminimal
torsion--matter coupling, with applications in a cosmological framework.
As it is usual in torsional formulation, one is based on the tetrad fields,
which  form an orthonormal
basis on the tangent space. In a coordinate basis, they are
$\mathbf{e}_A=e^\mu_A\partial_\mu $, and the metric is given by
\begin{equation}  \label{metricrel}
g_{\mu\nu}(x)=\eta_{AB}\, e^A_\mu (x)\, e^B_\nu (x),
\end{equation}
where  $x^\mu$ is the manifold point, $\eta_{AB}={\rm diag}
(1,-1,-1,-1)$), and
with Greek  and Latin indices   denoting  coordinate and
tangent indices respectively.
Additionally, one introduces the
Weitzenb\"{o}ck connection, given as
$\overset{\mathbf{w}}{\Gamma}^\lambda_{\nu\mu}\equiv e^\lambda_A\:
\partial_\mu
e^A_\nu$ \cite{Weitzenb23}, and therefore the   torsion tensor~reads
\begin{equation}
\label{torsten}
{T}^\lambda_{\:\mu\nu}\equiv\overset{\mathbf{w}}{\Gamma}^\lambda_{
\nu\mu}-%
\overset{\mathbf{w}}{\Gamma}^\lambda_{\mu\nu}
=e^\lambda_A\:(\partial_\mu
e^A_\nu-\partial_\nu e^A_\mu).
\end{equation}

Contraction of the torsion tensor gives the torsion scalar as
\begin{equation}
\label{torsiscal}
T\equiv\frac{1}{4}
T^{\rho \mu \nu}
T_{\rho \mu \nu}
+\frac{1}{2}T^{\rho \mu \nu }T_{\nu \mu\rho }
-T_{\rho \mu }^{\ \ \rho }T_{\
\ \ \nu }^{\nu \mu },
\end{equation}
which is used as the Lagrangian of the teleparallel theory.
Variation of the teleparallel action with respect to the tetrads
leads to  the same equations
with general relativity, and that is the reason that the corresponding theory is denoted as the teleparallel equivalent of general relativity (TEGR) \cite{Maluf:2013gaa,JGPereira,Maluf:1994ji}.

As mentioned in the Introduction, one can construct gravitational
modifications by extending  $T$ to
$T+f(T)$, writing  the action \cite{Cai:2015emx}
\begin{eqnarray}
\label{action0}
I = \frac{1}{16\pi G}\int d^4x \,e \left[T+f(T)+L_m\right],
\end{eqnarray}
where $e = \text{det}(e_{\mu}^A) = \sqrt{-g}$ and  $G$ is the gravitational
constant in units where the speed of light is set to 1. For completeness, in the
above action, we have added the total matter Lagrangian $L_m$.


One can now proceed to a further extension and allow for a nonminimal coupling
between torsion terms and the matter Lagrangian $L_m$.
Such  models have the theoretical advantage that  matter is
considered on  an equal footing with geometry and prove to lead to interesting
cosmological phenomenology \cite{Harko:2014sja}.
In this case, the action is written as
\begin{equation}
S= \frac{1}{16\pi G}\,\int
d^{4}x\,e\,\left\{T+f_{1}(T)+\left[1+\lambda\,f_{2}
(T)\right]\,L_{m}\right\},
\label{1}
\end{equation}
with $f_{i}(T)$ ($i=1,2$) two  arbitrary functions  $T$ and $\lambda$  a
coupling constant with dimensions of ${\rm
mass}^{-2}$. Variation with respect to the tetrad
leads to
\begingroup\makeatletter\def\f@size{9.5}\check@mathfonts
\def\maketag@@@#1{\hbox{\m@th\normalsize\normalfont#1}}%
\begin{eqnarray}
&&\left(1+f_{1}'+\lambda f_{2}' L_{m}\right) \left[e^{-1}
\partial_{\mu}{(e e^{\alpha}_{A} S_{\alpha}{}^{\rho \mu})}-e^{\alpha}_{A}
T^{\mu}{}_{\nu \alpha} S_{\mu}{}^{\nu\rho}\right]
+\left(f_{1}''+ \lambda f_{2}'' L_{m}
\right)  \partial_{\mu}{T} e^{\alpha}_{A} S_{\alpha}{}^{\rho\mu}
\nonumber\\
&&\;\; +e_{A}^{\rho} \left(\frac{f_{1}+T}{4}\right)
-\frac{1}{4} \lambda f_{2}' \,
\partial_{\mu}{T} e^{\alpha}_{A} \overset{\text{em}}{S}_{\alpha}{}^{\rho \mu}
+ \lambda f_{2}'\, e^{\alpha}_{A} S_{\alpha}{}^{\rho\mu} \,
\partial_{\mu}{L_{m}}
=4\pi G \left(1+\lambda f_{2}\right) e^{\alpha}_{A}
\overset{\text{em}}{T}_{\alpha}{}^{\rho}.
\label{eomsfull}
\end{eqnarray}\endgroup
%

In these field equations, primes denote derivatives with respect to $T$,
and  $\overset{\text{em}}{T}_{\rho}{}^{\nu}$  denotes   the total
matter  energy--momentum tensor. Moreover, we have defined
$
S_\rho^{\:\:\:\mu\nu}\equiv\frac{1}{2}\Big(K^{\mu\nu}_{\:\:\:\:\rho}
+\delta^\mu_\rho
\:T^{\alpha\nu}_{\:\:\:\:\alpha}-\delta^\nu_\rho\:
T^{\alpha\mu}_{\:\:\:\:\alpha}\Big)$, with
$K^{\mu\nu}_{\:\:\:\:\rho}\equiv-\frac{1}{2}\Big(T^{\mu\nu}_{
\:\:\:\:\rho}
-T^{\nu\mu}_{\:\:\:\:\rho}-T_{\rho}^{\:\:\:\:\mu\nu}\Big)$    the contorsion
tensor, while for convenience, we have introduced
\begin{equation}
\overset{\text{em}}{S}_{A}{}^{\rho
\mu}=\frac{\partial{L_{m}}}{\partial{\partial_{\mu}{e^{A}_{\rho}}}}.
\label{Stilde}
\end{equation}

In order to apply the above theory in a cosmological framework,  we adopt
the flat, homogeneous, and isotropic  Friedmann--Lema\^{i}tre--Robertson--Walker (FLRW) metric:
\begin{equation}
ds^2= dt^2-a^2(t)\,  \delta_{ij} dx^i dx^j,
\label{metr}
\end{equation}
which is obtained by the tetrad
$e_{\mu}^A={\rm
diag}(1,a,a,a)$,
where $a(t)$ is the scale factor.
Concerning the matter fluid, as usual, we choose
$\overset{\text{em}}{T}_{\mu\nu}=(\rho_{m}+p_{m}) u_{\mu} u_{\nu}-p_{m} g_{\mu
\nu}$, while for the
Lagrangian density, the natural and also efficient choice is
$L_{m}/(16 \pi G)=-\rho_{m}$ \cite{GroenHervik,BPHL,BPHL2}, which then gives
$\overset{\text{em}}{S}_{A}{}^{\rho \mu}=0$.  Alternative choices for the matter Lagrangian have also been suggested. For example, in \cite{AvAz2018}, it was shown that the on-shell Lagrangian of a perfect fluid depends on the microscopic properties of the fluid. Moreover, if the fluid is comprised of solitons, representing localized concentrations of energy with fixed rest mass and structure, then the average
on-shell Lagrangian of a perfect fluid can be obtained as $L_m = T_m=-\rho_m+3p_m$, where $T_m$ denotes the trace of the energy--momentum tensor. However, in the limit of dust, with $p<<\rho$, the matter Lagrangian $L_m=T_m$ reduces to $L_m=-\rho_m$.

Inserting the above considerations into the general field Equation
(\ref{eomsfull}),  we obtain the  Friedmann equations
\begin{equation}\label{H0}
H^{2}=\frac{8\pi G}{3} \left[1+\lambda
\left(f_{2}+12 H^{2} f_{2}' \right)\right] \rho_m-\frac{1}{6} \left(f_{1}+12
H^{2} f_{1}'\right),
\end{equation}
\begin{equation}\label{H00}
\dot{H}=-\frac{4\pi G\left( \rho _{m}+p_{m}\right) \left[ 1+\lambda
\left(
f_{2}+12H^{2}f_{2}^{\prime }\right) \right] }{1+f_{1}^{\prime
}-12H^{2}f_{1}^{\prime \prime }-16 \pi G \lambda \rho _{m}\left( f_{2}^{\prime
}-12H^{2}f_{2}^{\prime \prime }\right) }.
\end{equation}

As we observe, we obtain extra terms arising from both the torsional
modification, as well as from the nontrivial torsion--matter couplings.

The extended Friedmann equations  can be re-expressed  as
\begin{eqnarray}
3H^2&=& 8\pi G\left(\rho_{DE}+\rho_m  \right), \label{Fr1} \\
2\dot{H}+3H^2& =&-8\pi G\left(p_{DE}+p_m\right), \label{Fr2}
\end{eqnarray}
in which we have introduced an effective dark energy sector with
energy density and   pressure  given as
\begin{equation}
\label{rhode}
\rho_{DE}:=-\frac{1}{16 \pi G} \left(f_{1}+12 H^{2}
f_{1}'\right)+ \lambda \rho_m\left(f_{2}+12 H^{2} f_{2}' \right) ,
\end{equation}
\begin{eqnarray}\label{pde}
&&p_{DE}:= \left(\rho _m+p_m\right)
\left[\frac{  1+\lambda \left(
f_{2}+12H^{2}f_{2}^{\prime }\right)  }{1+f_{1}^{\prime
}-12H^{2}f_{1}^{\prime \prime }-16\pi G\lambda \rho _{m}\left( f_{2}^{\prime
}-12H^{2}f_{2}^{\prime \prime }\right) }-1\right]
\nonumber\\
&& \ \ \ \ \ \ \ \ \ \ \ \ \ +\frac{1}{16 \pi G} \left(f_{1}+12 H^{2}
f_{1}'\right)- \lambda \rho_m\left(f_{2}+12 H^{2} f_{2}' \right).
\end{eqnarray}

Using these, we can introduce   the equation-of-state parameter  for the
dark energy  sector through
\begin{eqnarray}
w_{DE}:= \frac{p_{DE}}{\rho_{DE}}.
\label{wDE}
\end{eqnarray}

To compare the theoretical predictions with observations, it is advantageous  to introduce the deceleration parameter $
q=\frac{d}{dt}\left(\frac{1}{H}\right)-1$,  and thus $q<0$ corresponds to
acceleration, while $q>0$ to deceleration.

\textls[-25]{As one can see, the energy densities and pressures satisfy
the total conservation~equation}
\begin{eqnarray}\label{cons}
\dot{\rho}_{DE}+\dot{\rho} _m +3H\left(\rho_{DE}+\rho
_m+p_{DE}+p_m\right)=0.
\end{eqnarray}

However, the crucial feature is that in such classes of theories, the matter and
dark energy sectors are not separately conserved. This lies in the basis of the
present work, as we will see below.

We close this section by  introducing for convenience the notations
\be
F_i=f_i+12H^2f^{\prime}_i, \qquad G_i=f_i^{\prime}-12H^2f_i^{\prime \prime}, \qquad i=1,2.
\ee

Then, from Equation~(\ref{Fr1}), we obtain the matter density as
\be\label{rhom}
\rho _m=\frac{3H^2+F_1/2}{8\pi G\left(1+\lambda F_2\right)}.
\ee

By assuming an equation of state of the form $p_m=(\gamma -1)\rho _m$, Equation~(\ref{H00}) gives
\bea\label{dotH}
\dot{H} =-\frac{\gamma \left(1+\lambda
F_2\right)\left(3H^2+F_1/2\right)}{\left(1+G_1\right)\left(1+\lambda
F_2\right)-2\lambda \left(3H^2+F_1/2\right)G_2}.
\eea

This is a first order ordinary differential equation, which fully determines the
cosmological evolution of the model once the functions $f_i$ ($i=1,2$) are
specified. In terms of the matter and dark energy densities, the deceleration
parameter can be written as
\be
q=\frac{1}{2}+\frac{p_m+p_{DE}}{\rho _m+p_{DE}}.
\ee

Finally, to facilitate the testing of the models with the observations, we
introduce the redshift $z$, defined as
$
1+z=\frac{1}{a}$.

Then, we can replace the derivatives with respect to the time with the derivatives with respect to $z$ according to the rule
$
\frac{d}{dt}=-(1+z)H(z)\frac{d}{dz}$.

\section{Irreversible Thermodynamics of Open Systems}\label{sect3}

In the present section, we briefly review the fundamentals of the thermodynamics
of irreversible processes in the presence of gravitationally induced matter
creation. The main motivation for considering this formalism is related
to the possibility of the interpretation of the nonconservation of the standard
matter energy--momentum tensor in theories with torsion--matter coupling, as
describing particle creation on a cosmological scale, a process that would
naturally require the use of the irreversible thermodynamics of open systems \cite{Pri0,Pri,Cal,Lima, Ha1, Ha2}.
Taking into account Equation~(\ref{cons}), it turns out that in the presence of a torsion--matter coupling, the energy--momentum balance equation contains extra terms,  as contrasted with the adiabatic conservation law of standard cosmology. From a cosmological point of view, these new terms can be interpreted as corresponding to dark energy. On the other hand, from the point of view of the thermodynamics of open systems, the same terms can be portrayed as corresponding to an irreversible matter creation process, describing the generation of particles from  torsion.
Particle production also acts as an
entropy source, which generates an effective entropy flux. Consequently, in the presence of the torsion--matter coupling,  the
temperature evolution of the thermodynamic systems is also modified.

\subsection{Second Law of Thermodynamics, Particle Generation Rates, and the Creation Pressure}

In the following, we define open systems  as specific thermodynamic systems that can transfer, via some dissipative processes,  both energy and matter
to their enclosing. In the matter--torsion coupling model, due to the nonconservation of the matter energy--momentum tensor, geometry  can  transfer matter and  energy to the cosmic background. On the other hand, in a closed thermodynamic system, only exchange
of energy (in the form of heat) occurs, and there is no exchange of matter with
its surroundings. A closed thermodynamic system is surrounded by walls that are immovable and rigid. Therefore,  the walls of a closed system cannot conduct
heat or perfectly reflect radiation, and hence they are impervious to particles and
nongravitational forces \cite{LaListat}.

As a second fundamental assumption, we adopt the physical idea that the cosmological evolution is irreversible. In the present context, this implies that torsion--matter coupling could generate physical particles (such as, for example, photons), but particles cannot directly generate geometric quantities, except via the gravitational field equations.

For a general relativistic fluid, its thermodynamic state can be characterized by using as basic macroscopic variables the
energy--momentum tensor $\overset{\text{em}}{T}_{\mu \nu}$, the entropy flux vector $s^{\mu}$, and the particle flux vector $N^{\mu}$.
For open thermodynamical systems, one must also admit the possibility of the modification of the matter content, due to irreversible particle production, or decay processes.

\subsubsection{Entropy Evolution in Irreversible Thermodynamics}

In the thermodynamical analysis of open systems, one generally assumes that the
entropy variation is the result of two distinct processes: the first is the entropy flow $d_eS$, while the second term gives
the entropy creation $d_iS$. Hence, in the presence of matter creation, the total entropy change $dS$ of an open
thermodynamic system can be written \mbox{as \cite{Pri0,Pri}}
\begin{equation}
dS = d_eS + d_iS,
\end{equation}
with the second law of thermodynamics requiring $d_iS > 0$. The total differential of the entropy is obtained as
\cite{Pri},
\begin{equation}\label{TdS}
\mathcal{T} d\left(sa^3\right)=d\left(\rho_m a^3\right)+p_m da^3-\mu
d\left(na^3\right),
\end{equation}
where by $\mathcal{T}$ we have denoted the temperature of the open thermodynamic system, $\mu $ is the chemical
potential, $n$ is the particle number density, while $s=S/a^3 $ denotes the entropy per unit volume, respectively. $\mu $ can be obtained from the relation
\begin{equation}
\mu n=\tilde{h}-\mathcal{T} s,
\end{equation}
where $\tilde{h}=\rho_m +p_m$ denotes the enthalpy of the system. Thus, both the entropy flow and the entropy production rate can be obtained from  Equation~(\ref{TdS}), giving the second law of thermodynamics.

For closed thermodynamic systems, and in the case of adiabatic
transformations, we have $dS=0$, and $d_iS=0$, respectively.  In any homogeneous and isotropic
cosmological model, the entropy flow term $d_eS$ identically vanishes, and thus $d_eS = 0$.

\subsubsection{Particle Production Rates, Creation Pressure, and Temperature Evolution}

For open thermodynamical systems, in which  irreversible matter creation/decay occurs, the changes of the particle numbers  must also be taken into account when considering the evolution laws. The particle creation processes can be included phenomenologically in the energy--momentum tensor, which must be  written as \cite{Pri0,Pri}
\be\label{tmunu}
\overset{\text{em}}{T}^{\mu \nu}=\left(\rho +p+p_c\right)u^{\mu }u^{\nu }-\left(p+p_c\right)g^{\mu \nu},
\ee
where $u^{\mu }$ is the four-velocity of the fluid, normalized according to $u_{\mu}u^{\mu}=1$, while  the creation  pressure $p_c$ characterizes matter creation, as well as other possible dissipative thermodynamic processes (note that this could hold in the case of soft cosmology, too \cite{Saridakis:2021qxb}). The total energy--momentum tensor $\overset{\text{em}}{T}_{\mu \nu}$  satisfies the usual covariant conservation law
\be\label{econs}
\nabla _{\nu }\overset{\text{em}}{T}^{\mu \nu}=0,
\ee
where by $\nabla _{\nu }$, we have denoted the covariant derivative with respect to the Riemannian metric $g_{\mu \nu}$ and to the corresponding Levi-Civita connection, respectively.

The particle flux vector $N^{\mu}$ is introduced via the definition $N^{\mu} =nu^{\mu }$,
and it obeys the general balance equation
\be\label{n}
\nabla _{\mu}N^{\mu }=\Psi \left(x^{\mu}\right) ,
\ee
where the function $\Psi  \left(x^{\mu}\right)$ gives the matter production rate. A positive $\Psi >0$ indicates a particle source, while a negative $\Psi <0$ corresponds to a particle sink. In standard general relativistic cosmology, usually  the particle production rate $\Psi $ is taken as zero, $\Psi \equiv 0$. The entropy flux $s^{\mu }$ is given by $s^{\mu }=n\sigma u^{\mu }$ \cite{Cal},
where $\sigma =s/n$ denotes the specific entropy per particle. From the second law of thermodynamics, it follows that $\nabla _{\mu }s^{\mu }\geq 0$. An important thermodynamical relation, the Gibbs equation, is given for an open thermodynamic system by \cite{Cal}
\be\label{gibbseq}
n\mathcal {T}d\sigma =d\rho -\frac{\rho +p}{n}dn.
\ee

We now derive the energy balance equation in an open thermodynamic system in the presence of particle creation. To achieve this goal, we multiply both sides of Equation~(\ref{econs}) by $u^{\mu}$, thus obtaining the relation
\begin{eqnarray}
u_{\mu }\nabla _{\nu }T^{\mu \nu } &=&u_{\mu }\left( \rho +p+p_{c}\right)
\nabla _{\nu }(u^{\mu }u^{\nu })-u_{\mu }\nabla ^{\mu }\left( p+p_{c}\right)
+u_{\mu }\nabla _{\nu }\left( \rho +p+p_{c}\right) u^{\mu }u^{\nu }
\nonumber \\
&=&\left( \rho +p+p_{c}\right) (u_{\mu }u^{\nu }\nabla _{\nu }u^{\mu
}+u_{\mu }u^{\mu }\nabla _{\nu }u^{\nu })+u^{\nu }\nabla _{\nu }(\rho
+p+p_{c})-\left( \dot{p}+\dot{p_{c}}\right)   \nonumber \\
&=&\dot{\rho}+\left( \rho +p+p_{c}\right) \nabla _{\nu }u^{\nu }=0,
\end{eqnarray}
where we have denoted $\dot{\rho}=u^{\mu}\nabla_{\mu}\rho$, and we have used the mathematical equalities $u_{\mu}u^{\mu}=1$  and $u_{\mu}u^{\nu}\nabla_{\nu}u^{\mu}=0$, respectively.  Therefore, for open systems in the presence of particle production, the energy conservation equation can be written in the general covariant form as
\begin{equation}\label{dotrho}
\dot{\rho}+(\rho+p+p_c)\nabla_{\mu}u^{\mu}=0.
\end{equation}

To obtain the entropy change, we use the relation (\ref{dotrho}) in the Gibbs Equation (\ref{gibbseq}), together with Equation~(\ref{n}). Thus, we successively obtain
\be
0=\dot{\rho}-\frac{\rho+p}{n}u^{\mu}\nabla_{\mu}n+
T\nabla_{\mu}s^{\mu}-T\sigma\nabla_{\mu}N^{\mu},
\ee
and
\be
nTu^{\mu}\nabla_{\mu}\sigma=-p_c\nabla_{\mu}u^{\mu}-\left( \frac{\rho+p}{n}-T\sigma\right) \nabla_{\mu}N^{\mu},
\ee
respectively.  Hence, for the entropy balance, we find the equation  \cite{Cal}

\begin{equation}\label{baentro}
\nabla_{\mu}s^{\mu}=-\frac{p_c\Theta}{T}-\frac{\mu\Psi}{T},
\end{equation}
where $\mu=(\rho+p)/n-T\sigma$ is the chemical potential, while $\Theta=\nabla_{\mu}u^{\mu}$ denotes the expansion of the fluid.

For  cosmological applications, we assume that matter is generated in thermal equilibrium with the particles previously
existing. This implies that the entropy increases due to particle generation processes only. As for the creation pressure  $p_c $, associated with new particle formation, in the following, we assume for it the phenomenological ansatz \cite{Pri,  Cal}
\be
p_c =-\alpha \left(x^{\mu}\right)\frac{\Psi}{\Theta},
\ee
where the function $\alpha \left(x^{\mu}\right)$ satisfies the condition $\alpha \left(x^{\mu}\right)>0$, $\forall x^{\mu}\in \mathbb{R}$. Hence, we obtain the entropy balance equation as
\bea
\nabla _{\mu }s^{\mu }&=&\frac{\Psi}{T}\left[\alpha \left(x^{\mu}\right) -\mu \right]=\Psi \sigma +\left[\alpha \left(x^{\mu}\right) -\frac{\rho +p}{n}\right]\frac{\Psi}{T}
\nonumber\\
&=&\Psi \sigma +n\dot{\sigma },
\eea
where  $\dot{\sigma}=u^{\mu }\nabla _{\mu }\sigma $.  With the use of Equation~(\ref{baentro}) for the specific entropy production $\dot{\sigma }$, we find the relation \cite{Cal}
\be\label{eq2}
\dot{\sigma }=\frac{\Psi}{nT}\left[\alpha \left(x^{\mu}\right) -\frac{\rho +p}{n}\right].
\ee

In the following, we narrow our thermodynamic formalism by setting the restriction that the specific entropy of the newly produced matter is a constant, $\sigma ={\rm constant}$. Under this assumption, from Equation~(\ref{eq2}), we obtain for $\alpha \left(x^{\mu}\right)$ the relation $\alpha \left(x^{\mu}\right)=\left(\rho +p\right)/n$.  Hence, it follows that the creation pressure generated by the irreversible matter production in open systems has the simple mathematical form \cite{Cal}
\be\label{pc0}
p_c=-\frac{\rho +p}{n\Theta }\Psi .
\ee

From  the condition of the constancy of $\sigma$, it follows that the Gibbs equation can be reformulated as
\be\label{ad}
\dot{\rho }=\left(\rho +p\right)\frac{\dot{n}}{n}.
\ee

\subsection{Cosmological Matter Generation}

We assume that the geometric properties of the Universe are described by the flat isotropic and homogeneous FLRW metric, given by   (\ref{metr}). In order to study the significance of particle  creation  at   cosmological scales, we assume that in a volume $V$, the universe contains $N$ ordinary baryonic  particles. We denote the energy density and the thermodynamic pressure of the matter system by $\rho _m$ and $p_m$, respectively. The second law of thermodynamics can be written down for such a system as \cite{Pri}
\begin{equation}
\frac{d}{dt}\left( \rho_m a^{3}\right) +p_m \frac{d}{dt}a^{3}=\frac{dQ}{dt}+\frac{%
\rho_m +p_m }{n}\frac{d}{dt}\left( na^{3}\right) ,  \label{21}
\end{equation}%
where by $dQ$, we have denoted the heat received by the system in a finite time $dt$, and $n=N/V$
is the particle number density. In the adopted FLRW metric (\ref{metr}), and by taking into account the cosmological principle, only adiabatic
transformations  $dQ=0$ are possible. Thus, we can ignore large scale heat transfer processes at the cosmological scale.
However, as can easily be seen from  Equation~(\ref{21}),  even under the premise of
adiabatic transformations,  in the second law of thermodynamics, as given by Equation~(\ref{21}), a term
$[(\rho_m +p_m )/n]d\left( na^{3}\right) /dt$ is still present.  This term takes into
account the temporal change of the cosmological particle number densities.

Therefore, in the formalism of the irreversible thermodynamics description of open systems, even in the case of adiabatic transformations $dQ=0$, a ``heat''-type term appears, which corresponds to the internal energy of the system. This term is due to the time variation in the particle number $n$. For adiabatic
transformations, $dQ/dt=0$, and Equation~(\ref{21}) takes the form
\begin{equation}
\dot{\rho}_m+3(\rho_m +p_m )H=\frac{\rho_m +p_m}{n}\left( \dot{n}+3Hn\right) .
\label{cons0}
\end{equation}

For the time variation of the particle number density we assume the balance equation
\begin{equation}
\dot{n}+3nH=\Psi n,  \label{22}
\end{equation}%
where  $\Psi $, the particle creation rate, is  a non-negative quantity. Therefore, the energy balance equation can be written in an
equivalent form as
\begin{equation}\label{41}
\dot{\rho}_m+3(\rho_m +p_m)H=(\rho_m +p_m)\Psi .
\end{equation}

For adiabatic transformations, Equation~(\ref{21}) can be reformulated as an effective energy conservation equation \cite{Pri},
\begin{equation}
\frac{d}{dt}\left( \rho_m a^{3}\right) +\left( p_m +p_{c}\right) \frac{d}{dt}%
a^{3}=0,
\end{equation}%
or, in an alternative representation, as
\begin{equation}
\dot{\rho}_m+3\left( \rho_m +p_m +p_{c}\right) H=0,  \label{comp}
\end{equation}%
where we have introduced the creation pressure $p_{c}$, given in the cosmological context \mbox{by \cite{Pri}}
\begin{eqnarray}  \label{pc1}
p_{c} &=&-\frac{\rho_m +p_m}{n}\frac{d\left( na^{3}\right) }{da^{3}}  \notag \\
&=&-\frac{\rho_m +p_m}{3nH}\left( \dot{n}+3nH\right) = -\frac{\rho_m +p_m}{3}\frac{%
\Psi }{H}.
\end{eqnarray}%

We now introduce  the entropy flux four-vector $S^{\mu }$, specified according to \cite{Cal}
\begin{equation}
S^{\mu }=n\sigma u^{\mu },
\end{equation}%
where $\sigma =S/N$ denotes, as usual,  the specific entropy per particle. Note that the entropy flux $S^{\mu }$ must
satisfy the second law of thermodynamics, which requires the restriction  $%
\nabla _{\mu }S^{\mu }\geq 0$. The Gibbs
relation 
and the definition of the chemical potential $\mu $ 
provides
\begin{eqnarray}
\nabla _{\mu }S^{\mu } &=&\Psi \frac{n}{T}\left( \frac{\tilde{h}}{n}-\mu \right)=\left( \dot{n}+3nH\right) \sigma +nu^{\mu }\nabla
_{\mu }\sigma  \notag  \label{48} \\
&=&\frac{1}{\mathcal{T}}\left( \dot{n}+3Hn\right) \left( \frac{\tilde{h}}{n}-\mu \right) ,
\end{eqnarray}%
where we have used the thermodynamic identity
\begin{equation}
n \mathcal{T} \dot{\sigma}=\dot{\rho}_m-\frac{\rho_m +p_m}{n}\dot{n}=0,
\end{equation}%
which is a consequence of Equation~(\ref{cons0}).

On the other hand, particle production contributes to the total entropy increase.
The time variation of $d_iS$ is obtained as \cite{Pri}
\begin{eqnarray}\label{53}
\mathcal{T}\frac{d_iS}{dt}&=&\mathcal{T}\frac{dS}{dt}=\frac{\tilde{h}}{n}\frac{d}{dt}\left(na^3\right)-%
\mu \frac{d}{dt}\left(na^3\right)  \notag \\
&=&\mathcal{T}\frac{s}{n}\frac{d}{dt}\left(na^3\right)\geq 0.
\end{eqnarray}

From Equation~(\ref{53}), we now obtain for the temporal change of the newly created entropy the relation
\begin{equation}\label{43}
\frac{dS}{dt}=\frac{S}{n}\left( \dot{n}+3Hn\right) =\Psi S\geq 0.
\end{equation}%

A realistic thermodynamic system is described by two basic thermodynamic
variables, the temperature $\mathcal{T}$ and the particle number density $n$,
respectively. In thermodynamic equilibrium, the energy density $%
\rho_m $ and the thermodynamic pressure $p_m$ of the matter are obtained generally, in terms of $n$ and $\mathcal{T}$, in a parametric form given by
\begin{equation}
\rho_m =\rho_m (n,\mathcal{T}), \qquad   p_m=p_m(n,\mathcal{T}).  \label{51}
\end{equation}%

Hence, the energy conservation Equation (\ref{41}) can be formulated as
\begin{equation}
\frac{\partial \rho_m }{\partial n}\dot{n}+\frac{\partial \rho_m }{\partial \mathcal{T}}%
\dot{\mathcal{T}}+3(\rho_m +p_m)H=\Psi n.
\end{equation}%

From the general thermodynamic relation \cite{Cal}
\begin{equation}
\frac{\partial \rho_m }{\partial n}=\frac{\tilde{h}}{n}-\frac{\mathcal{T}}{n}\frac{\partial p_m}{%
\partial \mathcal{T}},
\end{equation}%
it follows that the temperature variation of the newly created particles can be obtained as
\begin{equation}\label{54}
\frac{\dot{\mathcal{T}}}{\mathcal{T}}=c_{s}^{2}\frac{\dot{n}}{n}=c_{s}^{2}\left( \Psi
-3H\right) ,
\end{equation}%
where we have introduced the speed of sound $c_{s}$ in the cosmological matter, defined according to  $c_{s}^{2}=\partial
p_m/\partial \rho_m $.

If the newly created particles satisfy a barotropic
equation of state of the form $p_m =\left( \gamma -1\right) \rho_m $, $1\leq
\gamma \leq 2$, then it follows that the temperature $\mathcal{T}$ evolves according to the~relation
\begin{equation}
\mathcal{T}=\mathcal{T}_{0}n^{\gamma -1}.
\end{equation}

\subsection{Particle Creation and Bulk Viscosity}

An interesting physical interpretation of the matter production processes was initiated in \cite{Zeld} and further developed in
\cite{Mur,Hu}, respectively. The basic idea of this approach is the interpretation of the viscosity
of the cosmological fluid as corresponding to a phenomenological characterization of the production of particles in an
expanding Universe. Hence, the matter creation process can be equivalently described through the addition of
an effective bulk viscous type pressure into the energy--momentum tensor of the cosmological
matter. From a physical point of view, the viscous pressure can be interpreted in terms of the viscosity of the vacuum \cite{Zeld,Mur,Hu}. Moreover, in the general energy balance equation of a general relativistic fluid, any source term can be
reformulated as an effective bulk viscosity coefficient \cite{Mart,Mart1}.

In the presence of bulk viscosity, assumed to  represent the only dissipative process, for a general relativistic fluid, the energy--momentum tensor can be obtained as \cite{Mart, Mart1}
\begin{equation}
\overset{\text{em}}{T}_{\mu \nu}=\left(\rho_m +p_m+\Pi \right)u_{\mu}u_{\nu}-\left(p_m+\Pi\right) g_{\mu\nu},
\end{equation}
where by $\Pi $, we have denoted the bulk viscous pressure. The particle flow vector $N^{\mu}$
is defined in the usual way as  $N^{\mu}=nu^{\mu}$. On the other hand, in the causal formulation of thermodynamics,
the entropy flow vector $S^{\mu}$ is given by \cite{Isr1, Isr2,Isr3}
\begin{equation}\label{flow}
S^{\mu}=sN^{\mu}-\frac{\tau \Pi^2}{2\xi {\mathcal{T}}}u^{\mu},
\end{equation}
where by $\tau $, we have denoted the relaxation time, and $\xi $ is the bulk
viscosity coefficient. In  (\ref{flow}), only small,
second-order departures from equilibrium are included. In the FLRW geometry, for a cosmological fluid in a homogeneous and isotropic Universe in the presence of bulk viscous dissipative processes, the energy conservation equation is given by
\begin{equation}\label{comp1}
\dot{\rho}_m+3\left(\rho_m +p_m +\Pi\right)H=0.
\end{equation}

A simple comparison between Equation~(\ref{comp1}), giving the energy conservation equation for
a cosmological fluid in the presence of bulk viscosity, and Equation~(\ref{comp}),
describing the creation of particles in the framework of the irreversible thermodynamics of open systems, shows
that the two equations are mathematically equivalent  if
\begin{equation}
p_c=\Pi.
\end{equation}

Hence, matter production can also be described phenomenologically by introducing an effective, bulk viscous-type pressure in the
energy--momentum tensor of the cosmological matter content of the Universe, with the bulk viscous pressure $\Pi $ assuming the role of a creation pressure.

In order to further investigate the relation between bulk viscosity and particle creation, we assume that the newly
generated particles satisfy an equation of state of the form
\begin{equation} \label{9_1}
\rho_m =\rho_{0}\left( \frac{n}{n_{0}}\right) ^{\gamma }=kn^{\gamma },
\end{equation}%
with $\gamma $ the barotropic index, and where $n_{0}$ and $\rho_{0}$ are constants with  $%
k=\rho _{0}/n_{0}^{\gamma }$. By taking into account
Equation~(\ref{9_1}),  Equation~(\ref{comp1}) can be interpreted as  a particle conservation equation,
\begin{equation}
\dot{n}+3Hn=\Psi n,
\end{equation}%
where
\begin{equation}
\Psi =-\frac{\Pi }{\gamma H}
\end{equation}%
is the particle creation rate. $\Psi$ is  proportional to the bulk viscous pressure and inversely proportional to the Hubble function.
By using the equation of state (\ref{9_1}) in the Gibbs relation $\mathcal{T}d%
s=d(\rho_m /n)+p_m d(1/n)$, we obtain $s=s_{0}=\mathrm{constant}$,
a relation indicating that matter is created with a constant entropy density.

However, it is important to point out the existence of a major distinction between the two considered interpretations of matter creation.  The main difference appears in the expressions of entropy
production rates. The entropy production in the thermodynamical theory of open systems with particle
creation is given by \cite{Cal}
\begin{equation}\label{68}
\nabla _{\mu }S^{\mu }=-\frac{3Hp_c}{{\mathcal{T}}}\left(1+\frac{\mu \Psi n}{3Hp_c}%
\right)\geq 0,
\end{equation}
while in the causal thermodynamics, the entropy production
rate is \cite{Mart}
\begin{equation}\label{69}
\nabla _{\mu}S^{\mu}=-\frac{\Pi}{{\mathcal{T}}}\Bigg[3H+\frac{\tau }{\xi}\dot{\Pi}+%
\frac{\tau }{2\xi }\Pi\left(3H+\frac{\dot{\tau}}{\tau }-\frac{\dot {\xi}}{\xi%
}-\frac{\dot{{\mathcal{T}}}}{{\mathcal{T}}}\right)\Bigg].
\end{equation}

As one can see from Equations~(\ref{68}) and (\ref{69}), in the thermodynamics of open systems, the entropy
generation temporal rate is directly proportional to $p_c$, the creation pressure, while in the
causal thermodynamic description, $\nabla _{\mu
}S^{\mu }$ is quadratic term in the creation pressure, $\nabla _{\mu }S^{\mu
}\propto p_{c}^{2}/\xi {T}$. Moreover, a new dynamical
physical quantity, the bulk viscosity coefficient $\xi$, appears in the thermodynamic formalism.

\section{Cosmological Evolution and Particle Generation with Torsion--Matter Couplings}\label{sect4}

In the present section, we will investigate, from the point of view of the thermodynamics of irreversible processes, cosmological models with  the nonminimal torsion--matter coupling, given by Equation~(\ref{1}). Hence, as a starting point, we will interpret the energy balance Equation (\ref{cons}) as describing a process of particle creation, due to an effective energy transfer from torsional geometry/gravity to ordinary matter. Consequently, all the physical parameters describing particle creation also have a geometric origin.

We will also compare the cosmological predictions of the torsion--matter coupling gravity theory with the similar predictions of the $\Lambda$CDM model. To do so, we assume that the matter constituent of the late Universe is constituted of dust matter only, and thus we neglect the thermodynamic pressure of the cosmological matter. Then, the matter density varies according to
\be
\rho_m =\frac{\rho _0}{a^3}=\rho _0(1+z)^3,
\ee
where $\rho _0$ is the present day matter density. In the $\Lambda$CDM model,
the time evolution of the Hubble function is given by \cite{e8}
\be
H(z)=H_0\sqrt{\left(\Omega _b+\Omega _{DM}\right)(1+z)^{3}+\Omega _{\Lambda}},
\ee
where $H_0$ is the present day value of the Hubble function, and $\Omega _b$, $\Omega _{DM}$, and $\Omega _{\Lambda}$ represent
the density parameters of the baryonic matter, of the dark matter, and of the dark energy,
respectively. The density parameters satisfy the closure relation $\Omega _b+\Omega
_{DM}+\Omega _{\Lambda}=1$, which follows from the flatness of the Universe.

For the matter density parameters, we will use the numerical values $\Omega
_{DM}=0.259$, $\Omega _{b}=0.049$, and $\Omega _{\Lambda}=0.691$ \cite{1h}, respectively. The total matter
density parameter $\Omega _m=\Omega _{DM}+ \Omega _b$ is equal to
$\Omega _m=0.3089$. The present day value of the deceleration parameter, as predicted by the $\Lambda$CDM model, is $q(0)=-0.538$.

\subsection{Thermodynamic Parameters of Particle Creation}

A simple comparison between Equations~(\ref{cons}) and (\ref{41}) gives for the particle creation rate $\Psi$ the expression
\begin{equation}\label{Gamma}
\Psi =-\frac{1}{\rho_m +p_m}\left[\dot{\rho}_{DE}+3H(\rho_{DE} +p_{DE})
\right].
\end{equation}

By eliminating the time derivative of $\rho _{DE}$ with the help of the first Friedmann equation, we obtain for $\Psi$ the alternative expression
\be\label{Gamma1}
\Psi=-\frac{3H}{\rho _m+p_m}\left[\frac{\dot{H}}{4\pi G}-\frac{\dot{\rho}_m}{3}+\rho _{DE}+p_{DE}\right].
\ee

The requirement that the torsion--matter coupling acts as a particle source imposes the condition $\Psi>0$, which gives
\be
\rho _{DE}+p_{DE}<\frac{\dot{\rho}}{3}-\frac{\dot{H}}{4\pi G},
\ee
or, equivalently,
\be
\frac{\left(\rho _m+p_m\right)\left(1+\lambda F_2\right)}{1+G_1-16\pi G\lambda \rho _m G_2}<\rho_m+p_m+\frac{\dot{\rho}}{3}-\frac{\dot{H}}{4\pi G}.
\ee

Finally, the creation pressure can be obtained as
\begin{equation}\label{pc}
p_{c}=-\frac{\rho _m+p_m}{3}\frac{\Psi}{H}=\frac{1}{3H}\left[\dot{\rho}_{DE}+3H(\rho_{DE} +p_{DE})
\right].
\end{equation}

A positive particle creation rate $\Psi>0$ generates a negative creation pressure. By imposing a linear barotropic equation of state for the cosmological matter $p_m=\left(\gamma -1\right)\rho _m$, Equation~(\ref{41}) can be integrated to give for the variation of the matter energy density the expression
\be\label{rhom1}
\rho _m=\frac{\rho _{m0}e^{\gamma \int{\Psi \left(t'\right)dt'}}}{a^{3\gamma}},
\ee
where $\rho _{m0}$ is an arbitrary constant of integration. Similarly, for the temporal change of the particle number density, we obtain the relation
\be
n=\frac{n_0e^{ \int{\Psi \left(t'\right)dt'}}}{a^3},
\ee
with $n_0$ as a constant of integration.

Now, with the use of Equation~(\ref{43}),
for the entropy generated due to the particle creation induced by the nonminimal torsion--matter coupling,  we obtain the equation
\begin{equation}\label{entff}
\frac{1}{S}\frac{dS}{dt}=-\frac{1}{\rho_m +p_m}\left[\dot{\rho}_{DE}+3H(\rho_{DE} +p_{DE})
\right].
\end{equation}

Hence, the entropy increase due to new matter formation can be obtained generally from the expression
\begin{equation}\label{entcr}
S(t)=S_{0}e^{\int_{0}^{t}{\Psi \left( t^{\prime }\right) dt^{\prime }}},
\end{equation}
where $S_{0}=S(0)$ is an integration constant. Additionally, concerning the entropy flux, it can be computed from the equation
\begin{eqnarray}
\nabla _{\mu }S^{\mu }&=&\dot{S}^0+3HS^0=\Psi \frac{n}{\mathcal{T}}\left( \frac{\tilde{h}}{n}-\mu \right)\nonumber\\
&=&-\frac{n}{\mathcal{T}(\rho_m +p_m)}
\left[\dot{\rho}_{DE}+3H(\rho_{DE} +p_{DE})
\right] \left( \frac{\tilde{h}}{n}-\mu \right) .
\end{eqnarray}

Finally, the temperature change of the particles created from the torsion--matter transfer can be obtained as a function of $\Psi$, the particle creation rate, from Equation~(\ref{54}), and it is given~by
\be
\mathcal{T}=\mathcal{T}_0\frac{e^{c_s^2\int_{0}^{t}{\Psi \left( t^{\prime }\right) dt^{\prime }}}}{a^{3c_s^2}}.
\ee

\subsection{Specific Cosmological Models}

In the following, we will investigate, from the perspective of the irreversible thermodynamics of open systems, several cosmological models, obtained by fixing the functional forms of $f_1$ and $f_2$.

\subsubsection{The De Sitter Solution}

We will first look for a de Sitter-type solution of the basic field \mbox{Equations (\ref{rhom}) and (\ref{dotH}),} with $H=H_0={\rm constant}$.   Since the Hubble function is a constant, the right hand side of Equation~(\ref{dotH}) vanishes if $1+\lambda F_2=0$ or $3H^2+F_1/2=0$. The conditions $1+\lambda F_2=0$,  $3H^2+F_1/2\neq 0$, and $\forall t\geq 0$ would make the matter energy density (\ref{rhom}) infinite for all times. Thus, the de Sitter type solution corresponds to  $3H^2+F_1/2= 0$ and $1+\lambda F_2\neq 0$, respectively. However, with this choice, the matter energy-density vanishes identically. Therefore, a vacuum de Sitter phase in gravitational theories with torsion--matter coupling can be obtained if the condition
\be
\left.f_1(T)\right|_{T\rightarrow -6H^2}+12H^2\left.f_1^{\prime}(T)\right|_{T\rightarrow -6H^2}=0,
\ee
is satisfied. As an example, let us consider that $f_1$ has the functional form $f_1(T)=T_0T^n$, with $T_0$ and $n$ constants. Then, the existence of the de Sitter solution imposes the constraint
\be
3H_0^2+T_0\left(-6H_0^2\right)^n+12H_0^2nT_0\left(-6H_0^2\right)^{n-1}=0,
\ee
on the model parameters $H_0$, $n$, and $T_0$. It is important to mention that in this case, not only does the energy density of the ordinary matter vanish identically, but so, too, do the effective dark energy and dark pressure. This also implies that the particle creation rate is zero, and therefore, no matter is created during the exponential de Sitter inflation.

\subsubsection{Models with Fixed Form of the Creation Rate}

In the presence of torsion--matter coupling, by assuming that matter creation takes place, the particle production rate $\Psi$ is fully determined by the matter density, the torsion scalar, and the coupling functions $f_1(T)$ and $f_2(T)$ via Equations (\ref{Gamma}) and (\ref{Gamma1}), respectively. Hence, in order to obtain the particle creation rate, one must first specify the geometric characteristics of the model. However, an alternative approach is also possible, in which we assume from the beginning that the particle creation rate is a fixed function of the cosmological time. Of course, the consistency of this approach with the general expressions for the particle creation rate Equation (\ref{Gamma}) must be checked, and the conditions for the applicability of the model must be carefully investigated.  As an example of this approach, we consider the case in which $\Psi$ is a constant, an assumption that holds at least for small time intervals, $\Psi=\Psi_0={\rm constant}$. Then from Equation~(\ref{rhom1}), it follows that the matter density evolves in time according to the relation
\be\label{89}
\rho _m(t)=\rho _{m0}\frac{e^{\gamma \Psi_0t}}{a^{3\gamma}}.
\ee

Once the matter density is known, Equation~(\ref{rhom}) gives the expression of the Hubble function, which fully determines the cosmological dynamics. In order to further investigate the cosmological implications of the model, we consider the simple case with $f_1=\alpha T$ and $f_2=-\beta T$, with $\alpha>0$ and $\beta >0$ constants. Then, we have $F_1=6\alpha H^2$ and $F_2=-6\beta H^2$, respectively. Equation~(\ref{rhom}) becomes
\be\label{rhompsi0}
\rho _m=\frac{3\left(1+\alpha\right)H^2}{8\pi G\left(1-6\beta \lambda H^2\right)},
\ee
giving
\be
H^2=\frac{8\pi G\rho_m}{3}\frac{1}{1+\alpha +16\pi G\beta \lambda \rho _m}.
\ee

With the use of Equation~(\ref{89}), we obtain
\be
H^2=\frac{\dot{a}^2}{a^2}=\frac{8\pi G\rho _{m0}}{3}\frac{e^{\Psi_0t}}{(1+\alpha)a^3+16\pi G\beta \lambda \rho _{m0}e^{\Psi_0t}},
\ee
where we have assumed that the content of the Universe consists of pressureless dust with $\gamma =1$. In the limit $(1+\alpha)a^3 \ll 16\pi G\beta \lambda \rho _{m0}e^{\Psi_0t}$, that is, when the dynamics of the Universe is dominated by particle creation, we obtain
\be\label{H01}
H^2=\frac{1}{6\beta \lambda}={\rm constant},
\ee
indicating that matter creation generates a de Sitter type expansion. In the opposite limit $(1+\alpha)a^3 \gg 16\pi G\beta \lambda \rho _{m0}e^{\Psi_0t}$, we obtain
\be
a\dot{a}^2=\frac{8\pi G\rho_{m0}}{3(1+\alpha)}e^{\Psi_0t},
\ee
giving
\be
a=\left[\frac{4\pi G\rho_{m0}}{(1+\alpha)\Psi_0}e^{\Psi_0t}+a_0\right]^{2/3},
\ee
where $a_0$ is an arbitrary constant of integration. It is interesting to note that even for $a_0=0$, $a(0)=\left[4\pi G\rho _{m0}/(1+\alpha)\Psi_0\right]\neq 0$, that is, in the present model the Universe begins its expansion from a finite value of the scale factor. For $a_0=0$, the expansion is again of the de Sitter type, with a constant Hubble function, and with the deceleration parameter $q=-1$.

We will consider now the cosmological situations under which the conditions of the constancy of the particle creation rate may be valid. By taking into account the adopted simple linear forms of $f_1$ and $f_2$, and considering a dust Universe with $p_m=0$, we first obtain
\be
\rho_{DE}=-\frac{3 \alpha  H^2}{8 \pi  G}-6 \beta  \lambda  H^2\rho _m,
\ee
and
\be
p_{DE}=\rho_m \left(\frac{1-6 \beta  H^2 \lambda }{\alpha +16 \pi  \beta  G \lambda
\rho_m+1}+\frac{3 \alpha  H^2}{8 \pi  G}+6 \beta   \lambda H^2 \rho_m-1\right),
\ee
respectively. Hence for the particle creation rate we obtain the expression\begingroup\makeatletter\def\f@size{9}\check@mathfonts
\def\maketag@@@#1{\hbox{\m@th\normalsize\normalfont#1}}%
\bea
\Psi &=&-\frac{H\left( 1-6\beta \lambda H^{2}\right) }{1+\alpha H^{2}}\Bigg\{
\frac{3\left( 1+\alpha H^{2}\right) }{1-6\beta \lambda H^{2}}\Bigg[ \frac{%
9\beta \lambda H^{2}\left( 1+\alpha H^{2}\right) }{4\pi G\left( 1-6\beta
\lambda H^{2}\right) }+\frac{3\alpha H^{2}}{8\pi G}\nonumber\\
&&+\frac{\left( 1-6\beta
\lambda H^{2}\right) ^{2}}{1+\alpha +6\beta \lambda -6\beta \lambda H^{2}}-1%
\Bigg] -\frac{12\beta \lambda \left( 1+\alpha H^{2}\right) \dot{H}}{%
1-6\beta \lambda H^{2}}-2\alpha \dot{H}-\frac{18\beta \lambda \left( \alpha
H^{4}+H^{2}\right) }{1-6\beta \lambda H^{2}}\nonumber\\
&&-3\alpha H^{2}\Bigg\} ,
\eea\endgroup
where we have also used Equation~(\ref{rhompsi0}). In the case of the de Sitter type expansion, with $H=H_0={\rm constant}$, the particle creation rate is a constant, and it can be approximated as
\be
\Psi _{0}\left( \alpha ,\beta ,\lambda ,H_{0}\right) \approx  3\left[ 1+%
\frac{4\left( \alpha +6\beta \lambda \right) }{8\pi G}H_{0}^{2}+\frac{1}{%
\alpha +6\beta \lambda }\right]  H_{0},
\ee
where we assumed that $6\beta \lambda H_{0}^{2}<<1$ and $\alpha H_{0}^{2}<<1$. If the Hubble function is given by the expression (\ref{H01}), then the particle creation rate becomes
\begin{equation}
\Psi _{0}\left( \alpha ,\beta ,\lambda \right) =\sqrt{\frac{3}{2}}\left[ 1+%
\frac{4\left( \alpha +6\beta \lambda \right) }{48\pi G\beta \lambda }+\frac{1%
}{\alpha +6\beta \lambda }\right] \frac{1}{\sqrt{\beta \lambda }}.
\end{equation}

This approach can easily be generalized to arbitrary time dependencies of the creation rate $\Psi (t)$ but implies the consistency check of the considered model.

\subsubsection{Models with Quadratic Torsion Dependence}

We consider now the case in which the functions $f_1(T)$ and $f_2(T)$ have a quadratic dependence on the torsion, and they are given by
\be
f_1(T)=-\Lambda +\alpha _1T^2, \qquad f_2(T)=\beta _1T^2.
\ee

In terms of the Hubble function, $f_1$ and $f_2$ are represented by the expressions \cite{Harko:2014sja}
\be
f_1(H)=-\Lambda +\alpha H^4, \qquad f_2(H)=\beta H^4,
\ee
where $\alpha =36\alpha _1$ and $\beta =36\beta _1$, respectively. Moreover, we immediately obtain $f_1'(H)=-\alpha H^2/3$, $f_2'(H)=-\beta H^2/3$, $f_1''(H)=\alpha /18$, and $f_2''(H)=\beta /18$, respectively. Then the field Equations (\ref{rhom}) and (\ref{dotH}) take the form
\be\label{dem0}
\rho _m(t)=\frac{3\alpha H^4-6H^2+\Lambda}{16\pi G\left(3\beta \lambda H^4-1\right)},
\ee
and
\be\label{dem1}
\dot{H}(t)=\frac{\left(3 \alpha  H^4-6 H^2+\Lambda \right) \left(3 \beta  \lambda
H^4-1\right)}{4 H^2 \left(\alpha +\beta  \lambda  \Lambda -3 \beta  \lambda
H^2\right)-4},
\ee\\
respectively. We now rescale the geometric and physical parameters according to the transformations
%
\bea
H=H_0h, \quad \tau =H_0t,  \quad \rho _m=\frac{3H_0^2}{8\pi G}r_m,
\quad
\alpha =\frac{\tilde{\alpha}}{H_0^2},  \quad
\beta =\frac{\tilde{\beta}}{H_0^2}, \quad \lambda =\frac{\tilde{\lambda}}{H_0^2}, \quad
\Lambda =3H_0^2\tilde{\Lambda},
\eea
%
%
\noindent where $H_0$ is the present day value of the Hubble function. Then, for the field equations, we obtain the dimensionless forms
\be
r_m(\tau)=\frac{\tilde{\alpha}h^4-2h^2+\tilde{\Lambda}}{2\left(3\tilde{\beta}\tilde{\lambda}h^4-1\right)},
\ee
and
\begin{equation}\label{94}
\frac{dh}{d\tau }=\frac{3}{4}\frac{\left( \tilde{\alpha}h^{4}-2h^{2}+\tilde{%
\Lambda}\right) \left( 3\tilde{\beta}\tilde{\lambda}h^{4}-1\right) }{\left(
\tilde{\alpha}+3\tilde{\beta}\tilde{\lambda}\tilde{\Lambda}-3\tilde{\beta}%
\tilde{\lambda}h^{2}\right) h^{2}-1},
\end{equation}
respectively. In terms of the redshift, Equation~(\ref{94}) becomes
\be
-(1+z)h\frac{dh}{dz }=\frac{3}{4}\frac{\left( \tilde{\alpha}h^{4}-2h^{2}+\tilde{%
\Lambda}\right) \left( 3\tilde{\beta}\tilde{\lambda}h^{4}-1\right) }{\left(
\tilde{\alpha}+3\tilde{\beta}\tilde{\lambda}\tilde{\Lambda}-3\tilde{\beta}%
\tilde{\lambda}h^{2}\right) h^{2}-1},
\ee
while for the redshift dependence of the deceleration parameter, we find the expression
\be
q=(1+z)\frac{1}{h}\frac{dh}{dz}-1.
\ee

\begin{figure}[H]
\includegraphics[scale=0.71]{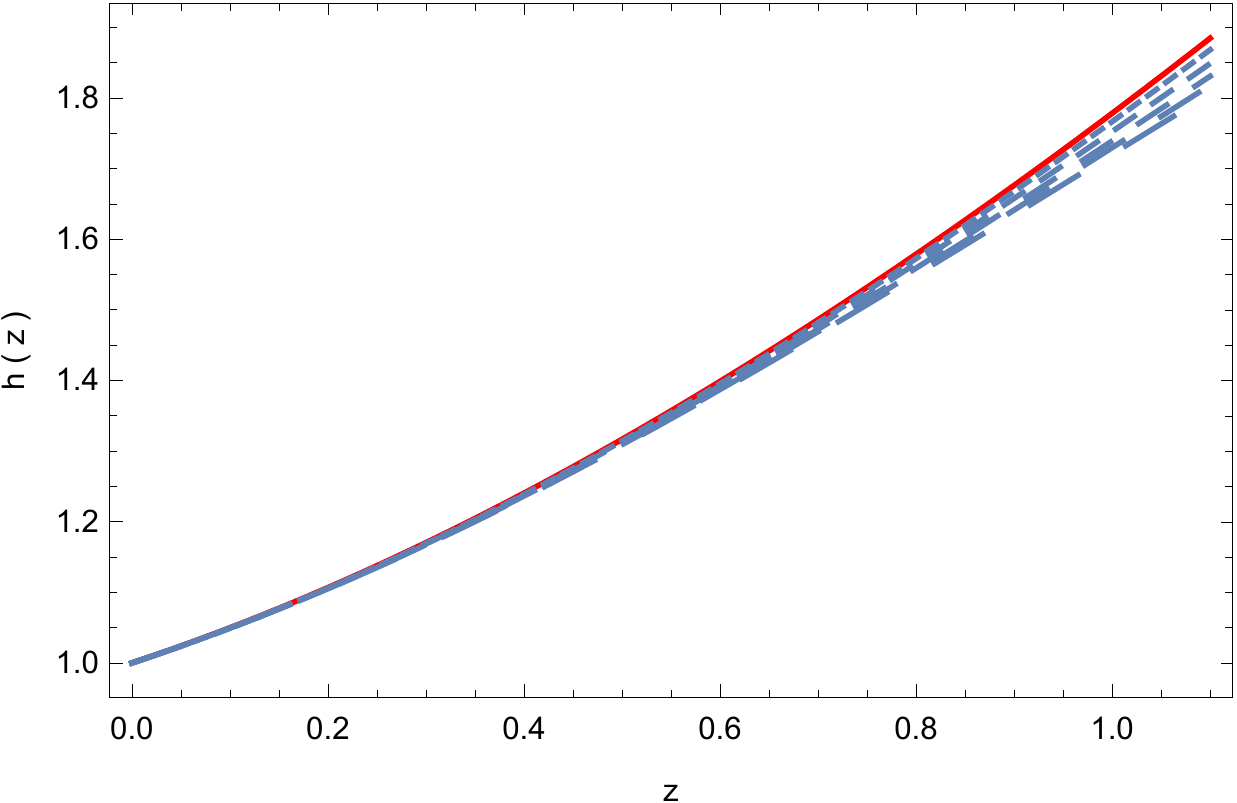}\\
\vspace{0.5cm}
\includegraphics[scale=0.71]{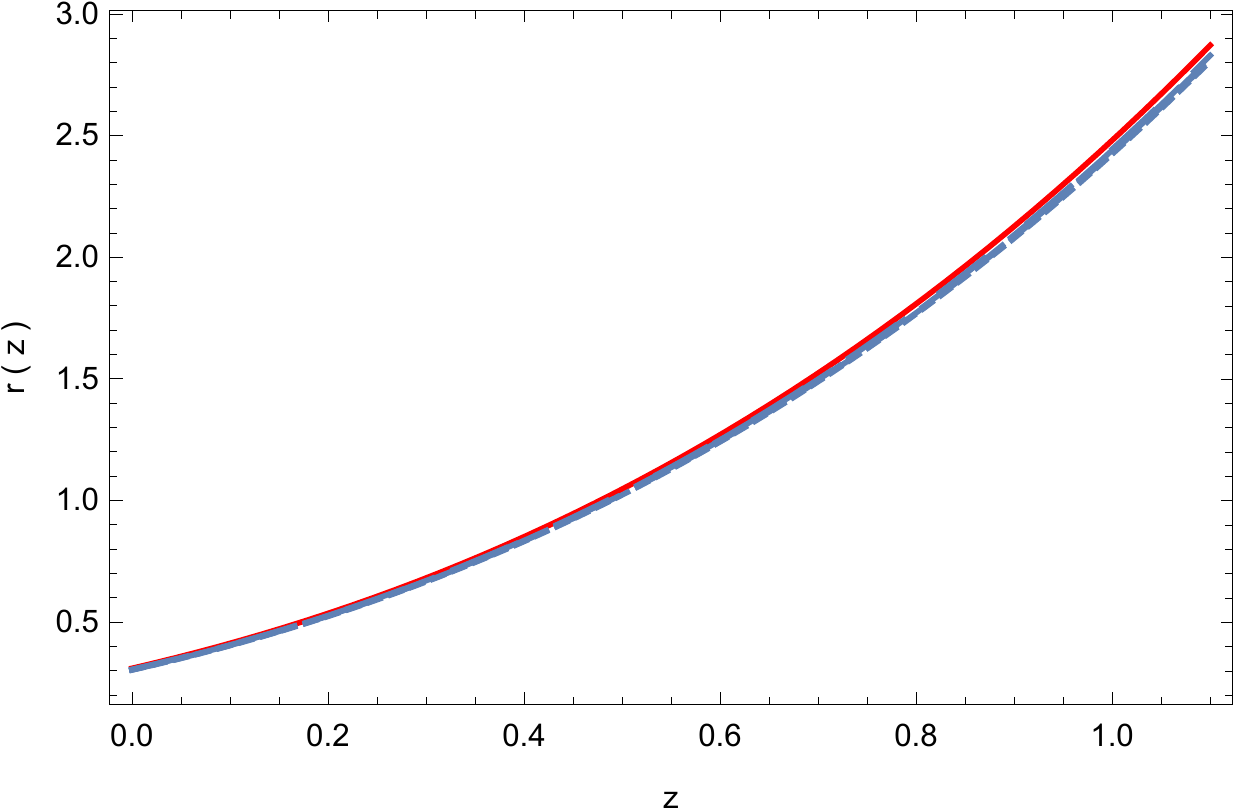}
\caption{{{The evolution of the dimensionless Hubble function $h(z)$ (top panel) and of the dimensionless matter density $r(z)$ (bottom panel) as a function of the redshift $z$ in the quadratic torsion--matter coupling model, for $\tilde{\alpha}=0.016$, $\tilde{\beta}=0.01$, $\tilde{\Lambda}=1.38$, and different values of the torsion--matter coupling constant $\tilde{\lambda}$: $\tilde{\lambda}=0.10$ (dotted curve), $\tilde{\lambda}=0.18$ (short-dashed curve), $\tilde{\lambda}=0.26$ (dashed curve),  $\tilde{\lambda}=0.32$ (long-dashed curve), and $\tilde{\lambda}=0.38$ (ultra-long-dashed curve), respectively.  The red solid line corresponds to the evolution of the Hubble function and of the matter density in the $\Lambda$CDM~model.}}}\label{f1}
\end{figure}

\begin{figure}[H]
\includegraphics[scale=0.71]{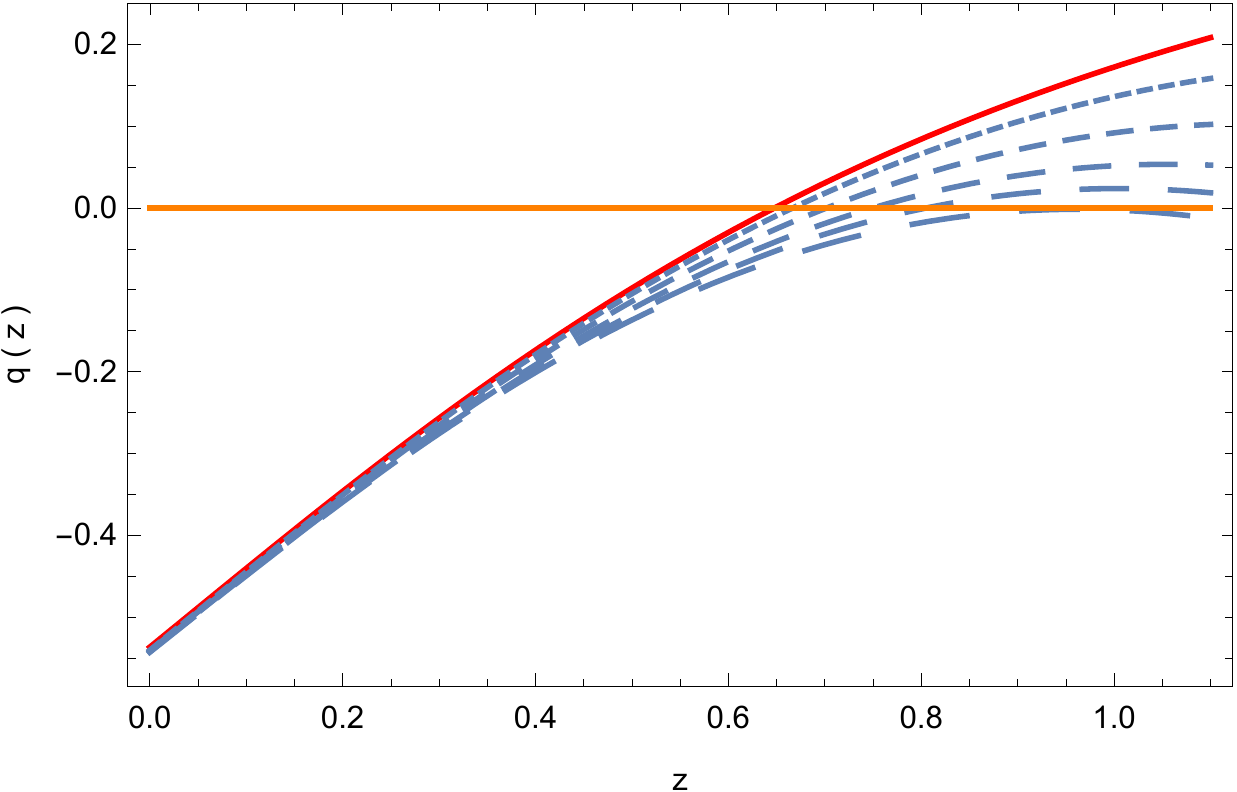}
\caption{{{The evolution of the deceleration parameter $q(z)$ as a function of the redshift $z$, in the quadratic torsion--matter coupling model, for $\tilde{\alpha}=0.016$, $\tilde{\beta}=0.01$, $\tilde{\Lambda}=1.38$, and different values of the torsion--matter coupling constant $\tilde{\lambda}$: $\tilde{\lambda}=0.10$ (dotted curve), $\tilde{\lambda}=0.18$ (short-dashed curve), $\tilde{\lambda}=0.26$ (dashed curve),   $\tilde{\lambda}=0.32$ (long-dashed curve), and $\tilde{\lambda}=0.38$ (ultra-long-dashed curve), respectively.  The red solid line describes evolution of the deceleration parameter in the $\Lambda$CDM model, while the horizontal line at $q=0$ is drawn for convenience.}}}\label{f2}
\end{figure}

The variations with respect to the redshift of the dimensionless Hubble function $h(z)$, of the matter density $r(z)$, and of the deceleration parameter $q(z)$ are represented in \mbox{Figures~\ref{f1} and \ref{f2},} respectively. For the sake of comparison, the predictions of the standard $\Lambda$CDM model are also represented.

As one can see from the top panel of Figure~\ref{f1}, presenting the evolution of $h(z)$, for the adopted set of the model parameters, the quadratic torsion--matter coupling model can reproduce the $\Lambda$CDM model up to a redshift of $z\approx 1$. For low redshift values, the behavior of the $h(z)$ model is almost independent of the numerical values of the coupling constant $\tilde{\lambda}$, but at higher redshifts, the evolution of $h$ is influenced by the variations of $\tilde{\lambda}$,  and significant differences between the $\Lambda$CDM and the present models appear. On the other hand, the matter density evolution, represented in the bottom panel of Figure~\ref{f1}, is almost identical with the $\Lambda$CDM evolution and essentially independent of the variation of $\tilde{\lambda}$. However, the behavior of the deceleration parameter $q(z)$, presented in Figure~\ref{f2}, indicates a strong dependency on the numerical values of $\tilde{\lambda}$, with important deviations from the $\Lambda$CDM model already appearing at redshifts higher than $z=0.5$. Furthermore, the value of the deceleration parameter at the transition point from the decelerating to the accelerating phase is also strongly dependent on the model parameters, and hence, the numerical value of the transition redshift $z_{tr}$ will impose  strong constraints on the gravitational  theories with torsion--matter coupling.

\paragraph{Matter Creation in the Early Universe}
In the very early universe, the Hubble function takes very large values. Hence, we can assume that at very large redshifts, the conditions $3\beta \lambda H^4 \gg 1$ and $3\alpha H^4 \gg -6H^2+\Lambda$ are valid. Moreover, we assume that at high redshifts, the Universe can be described as a radiation fluid, with the matter satisfying the equation of state $p_m=\rho_m/3$.  Then, for the matter energy-density, we obtain the expression
\be
\rho_m\approx \frac{\alpha}{16\pi G\beta \lambda}={\rm constant}.
\ee

Hence, from Equation~(\ref{41}), we obtain the following particle creation rate:
\be
\Psi\approx 3H.
\ee
\newpage
In the same approximation from Equation~(\ref{dem1}), we obtain the differential equation
\be
\dot{H}=-\frac{3}{4}\alpha H^4,
\ee
with the general solution satisfying the initial condition $H\left(t_0\right)=H_0$ given by
\be
H(t)=\frac{H_0}{\left[1+\left(9\alpha H_0^3/4\right) \left(t-t_0\right)\right]^{1/3}}.
\ee

Therefore, the time evolution of the scale factor   can then be obtained as
\be
a(t)=\exp\left\{\frac{1}{6\alpha H_0^2}\left[1+\frac{9 \alpha H_0^3}{4} \left(t-t_0\right)\right]^{2/3}  \right\},
\ee
while the deceleration parameter becomes
\be
q(t)=\frac{3 \alpha H_0^2 }{4\left[1+9 \alpha H_0^3 \left(t-t_0\right)/4\right]^{2/3}}-1.
\ee

Note  that in this approximation, the dynamics of the early universe is controlled by the parameter $\alpha$ only. If for $t=t_0$, the condition $3\alpha H_0^24>1$ holds, the Universe begins its evolution in a decelerating phase.  For the particle creation rate we find
\be
\Psi=\frac{3H_0}{\left[1+\left(9\alpha H_0^3/4\right) \left(t-t_0\right)\right]^{1/3}},
\ee
and the creation pressure is given by
\be
p_c=-\frac{\rho _m+p_m}{3}\frac{\Psi}{H}=-\frac{\alpha}{12\pi G\beta \lambda}.
\ee

Finally, for the entropy production we obtain
\be
S=S_0e^{\int{\Psi(t)dt}}=S_0e^{3\int{H(t)dt}}=S_0a^3,
\ee
or
\be
S=S_0\exp\left\{\frac{1}{2\alpha H_0^2}\left[1+\frac{9 \alpha H_0^3}{4} \left(t-t_0\right)\right]^{2/3}  \right\}.
\ee

Hence, a large amount of comoving entropy is created during this evolutionary phase, in which particle creation exactly compensates the decrease in the particle number density due to the expansion of the Universe.

\paragraph{Particle production in the late Universe} In the late Universe, the matter content can be described as dust with the equation of state $p_m=0$. Moreover, $H(t)$ takes much smaller values than in the early Universe. Hence, during this evolutionary phase, by assuming $3\beta \lambda H^4 \ll 1$ and $3\alpha H^4 \ll -6H^2+\Lambda$,  the matter energy-density (\ref{dem0}) can be approximated as
\be
\rho _m(t)=\frac{3 H(t)^2}{8\pi G}-\frac{\Lambda }{16\pi G}.
\ee

Therefore, for the particle creation rate, we obtain
\be\label{116}
\Psi=\frac{6H\dot{H}}{3H^2-\Lambda ^2}+3H.
\ee

In the limit of small $H(t)$, Equation~(\ref{dem1}) can be approximated as
\be
\dot{H}=\frac{1}{4} \left(\alpha  \Lambda +\beta  \lambda  \Lambda
^2-6\right)H^2 +\frac{\Lambda }{4},
\ee
and thus, integrating the above equation with the initial condition $H\left(t_0\right)=H_0$, we obtain for $H(t)$ the expression
\be
H(t)=\delta \tan \left[ \tan ^{-1}\left(
\frac{H_{0}}{\delta}\right) +\frac{%
\Lambda }{4\delta}\left( t-t_{0}\right) \right],
\label{Happrox1}
\ee
where we have denoted $\delta=\sqrt{\Lambda /\left[\Lambda (\alpha +\beta \lambda \Lambda )-6%
\right]}$.  Thus, for the particle creation rate, we obtain
\be
\Psi = \frac{3 \delta  \left(6 \delta ^2+\Lambda \right) \tan ^3\left[\tan
^{-1}\left(\frac{H_0}{\delta }\right)+\frac{\Lambda  (t-t_0)}{4 \delta
}\right]}{6 \delta ^2 \tan ^2\left[\tan ^{-1}\left(\frac{H_0}{\delta
}\right)+\frac{\Lambda  (t-t_0)}{4 \delta }\right]-\Lambda}.
\ee

By introducing the dimensionless variables $\epsilon H_0/\delta$, $\sigma =\delta ^2/\Lambda$, and $t=\tau /H_0$, the particle creation rate takes the form
\begin{equation}
\Psi (\tau )=3H_{0}\frac{6\sigma +1}{\epsilon }\frac{\tan ^{3}\left( \tan
^{-1}\epsilon +\frac{1}{4\epsilon \sigma }\tau \right) }{\sigma \tan
^{2}\left( \tan ^{-1}\epsilon +\frac{1}{4\epsilon \sigma }\tau \right) -1}.
\end{equation}

\begin{figure}[H]
\includegraphics[scale=0.71]{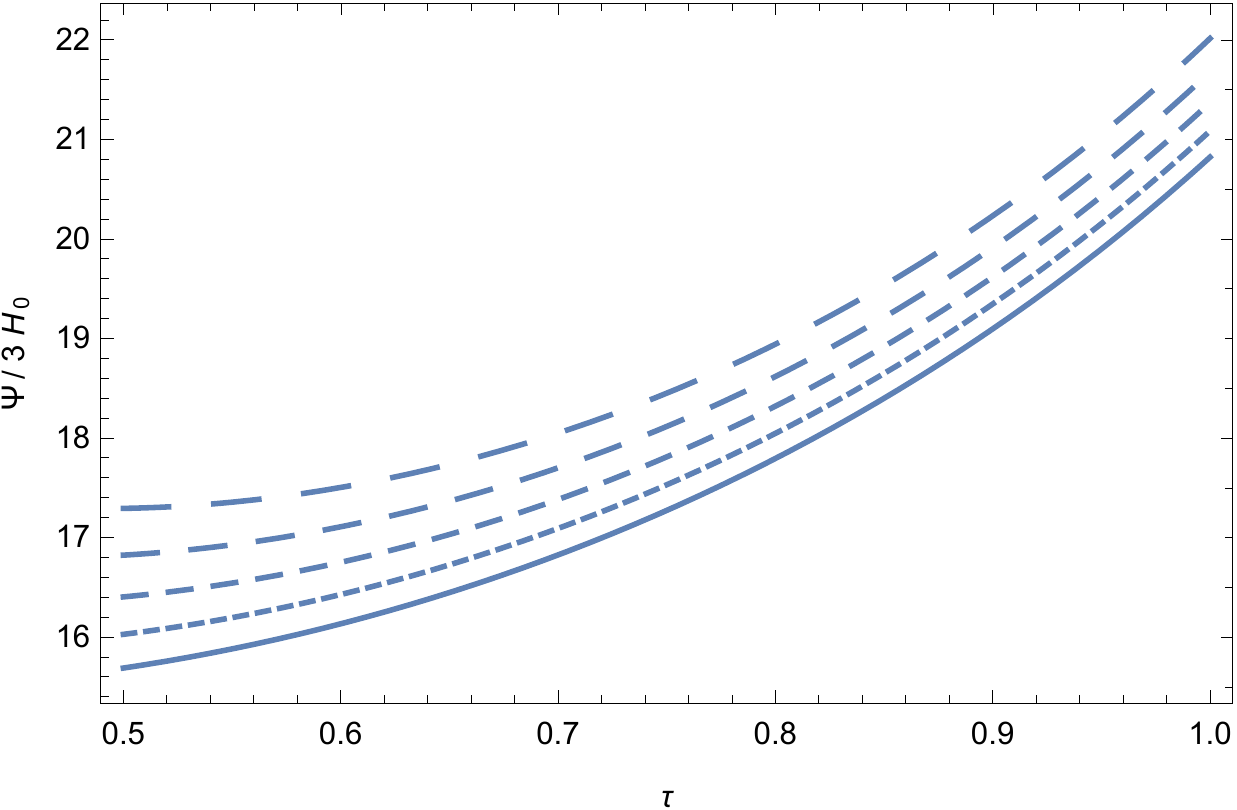}
\caption{{{Time dependence of the particle creation rate $\Psi/3H_0$ in the quadratic torsion coupling model in the late Universe for $\sigma =0.5$ and different values of $\epsilon$: $\epsilon =1.9$ (solid curve), $\epsilon =1.85$ (dotted curve), $epsilon =1.80$ (short dashed curve), $\epsilon =1.75$ (dashed curve), and $\epsilon =1.70$ (long dashed curve), respectively.}}}\label{f4}
\end{figure}

The variation of the ratio $\Psi/3H_0$ is represented in Figure~\ref{f4}. As one can see from the Figure, the particle creation rate is a monotonically increasing function of time, and its numerical values strongly depend on the model parameters $\epsilon $ and $\sigma$. For the considered range of parameters, one can estimate a present day value of the particle creation rate of the order of $\Psi \approx 60H_0=1.31\times 10^{-16}\;{\rm s}^{-1}$.

For the matter energy density we find
\bea
\rho _m(t)=\frac{3\delta ^2}{8\pi G} \tan ^2\left[\tan
^{-1}\left(\frac{H_0}{\delta}\right)+\frac{\Lambda  \left(t-t_0\right)}{4
\delta}\right] -\frac{\Lambda }{16\pi G}.
\eea

The creation pressure is given by
\be
p_c=-\frac{\left(6 \delta ^2+\Lambda \right) \tan ^2\left[\tan
^{-1}\left(\frac{H_0}{\delta }\right)+\frac{\Lambda  (t-t_0)}{4 \delta
}\right]}{16 \pi  G},
\ee
and Equation~(\ref{entcr}) yields for the entropy creation the expression
\bea
S&=&S_0\cos ^{-\frac{2 \left(6 \delta ^2+\Lambda \right)}{\Lambda }}\left[\tan
^{-1}\left(\frac{H_0}{\delta }\right)+\frac{\Lambda  (t-t_0)}{4 \delta
}\right]\times
\nonumber\\
&&
\left\{-6 \delta ^2+\left(6 \delta ^2+\Lambda \right) \cos \left[2 \tan
^{-1}\left(\frac{H_0}{\delta }\right)+\frac{\Lambda  (t-t_0)}{2 \delta
}\right] +\Lambda \right\}.
\eea

Lastly, for the scale factor and for the deceleration parameter, we have the approximate representations
\be
a(t)=a_{0}\cos ^{-\frac{4\delta ^2}{\Lambda }}\left[ \tan ^{-1}\left(
\frac{H_{0}%
}{\delta}\right) +\frac{\Lambda }{4\delta}\left( t-t_{0}\right) %
\right],
\ee
where $a_0$ is an arbitrary constant of integration to be determined from the initial conditions, and
\be
q(t)=-\frac{\Lambda  \csc ^2\left[\tan
^{-1}\left(\frac{H_0}{\delta}\right)+\frac{\Lambda  \left(t-t_0\right)}{4
\sqrt{H_1}}\right]}{4 \delta ^2}-1,
\ee
respectively. Since $q<0$ for all $t\geq t_0$, the Universe is in an accelerated state of expansion, triggered by the high rate of particle creation.

\section{Discussion and Conclusions}\label{sect5}

The nonconservation of the energy--momentum tensor of the cosmological matter due to geometry--matter coupling represents an interesting, and at the same time intriguing, aspect of modified gravity, particularly in theories with a torsion--matter coupling. In the present work, we have explored the significance of the nonvanishing divergence of the matter energy--momentum tensor by adopting the theoretical perspective of the thermodynamics of irreversible processes, as introduced and developed in \cite{Pri0, Pri,Cal, Lima}. Our basic assumption is that  in modified gravity theories  with a geometry--matter coupling, there is an effective relocation of energy from gravity (geometry) to matter \cite{Harko:2014pqa}, leading to a temporal variation in the particle number density at the cosmological level. Thus, through particle creation, geometry (torsion) also acts as a source of entropy and internal energy, respectively. Particle production takes place in an expanding Universe, and there is a strong correlation between the cosmological dynamics and irreversible thermodynamics, since large levels of particle production could trigger the transition from deceleration to~expansion.

In this  paper, we have investigated, from a thermodynamical perspective, a particular model of the modified $f\left(T,L_m\right)$ gravity theory, with the action given by Equation~(\ref{1}). This model involves a coupling between an arbitrary function of torsion $f_2(T)$ and the matter Lagrangian. A particularly appealing consequence of the theory is the nonconservation of $\overset{\text{em}}{T}_{\mu \nu}$. We have interpreted this nonconservation as corresponding to a particle creation process, the energy source for this process being the spacetime torsion. By using the theoretical approach of the thermodynamics of open systems, we have obtained the basic physical quantities (particle creation rate, creation pressure, entropy) in terms of geometric quantities, which are constructed from the torsion scalar. In particular, the particle creation rate, given by Equation~(\ref{Gamma}), is determined by the effective dark energy and dark pressure terms in the generalized Friedmann equations, and hence, $\rho_{DE}$ and $p_{DE}$ can be interpreted thermodynamically as describing particle generation from geometry; thus, they have a dual cosmological meaning. They also generate an effective creation pressure (\ref{pc}), which is, from a physical point of view, a direct consequence of the presence of particle production from torsional geometry.

After developing the general thermodynamical understanding of modified gravity theories with torsion--matter coupling, we also investigated several particular cosmological models. The generalized gravitational field equations admit a de Sitter type solution for a vacuum Universe, with the particle creation fully suppressed. The vacuum state is not changed during the exponential expansion. Models with a fixed form of the particle creation rate lead to nonsingular cosmological evolutions, with the scale factor taking an initial finite value.

We have also analyzed in detail a particular cosmological model, in which the functions $f_1$ and $f_2$ are quadratic in the torsion scalar $T$, containing a functional dependence of the form $T^2$. The model gives a very good description of the standard $\Lambda$CDM model up to a redshift $z\approx 1$. However, significant differences  with respect to $\Lambda$CDM appear at higher redshifts. The thermodynamic description of the model can be studied analytically in the approximations of large $H(t)$ (large redshifts) and small $H(t)$ (low redshifts), respectively. These approximations may each be valid for small cosmological time intervals. While in the high redshift limit, the thermodynamic description is rather simple, with the cosmological matter density kept constant due to particle creation, and with constant creation pressure, in the low redshift limit, the evolution of the geometric and thermodynamical parameters can be expressed in terms of trigonometric functions, leading to an oscillatory long-term behavior. However, the period of the oscillations is greater than the actual age of the Universe, and hence, once we restrict our analysis to the time interval $\left(0,1/H_0\right)$, the long-term oscillatory features of the cosmological dynamics are not detectable. In our approach, we have fixed the numerical values of the model parameters to obtain an equivalence with the $\Lambda$CDM model at a qualitative/semi-quantitative level. In order to better fix the values of the parameters, and to estimate the range of equivalence with $\Lambda$CDM, a detailed fitting of the observational results is necessary.

In the thermodynamic formalism of open systems, the entropy creation term is defined via the positive particle creation rate $\Psi$,  so that $\dot{S}/S=\Psi \geq 0$. Hence, this definition indicates that in an expanding Universe with matter generation, the entropy of the matter $S$ will increase forever. However, according to the second law of thermodynamics,  all natural systems must approach a state of thermodynamic equilibrium, so that the entropy of the equilibrium system never decreases, thus satisfying the condition $\dot{S}\geq 0$. Moreover, $S$ must be concave when it approaches the equilibrium state, and thus it should satisfy the condition $\ddot{S}\leq  0$ \cite{Mim, E2}. However, the present model does not automatically satisfy these requirements. However, as discussed in detail in \cite{Mim,E2}, if one considers  that the total entropy of the Universe is the sum of the entropies of the apparent horizon and of the matter and radiation components inside it,  it can be shown that the entropy grows and is simultaneously concave. 
Hence, it turns that the second law of thermodynamics may still be valid in the case of an indefinite cosmological growth in the presence of particle creation. The conditions $\dot{S}\geq 0$ and $\ddot{S}=\left(\dot{\Psi}+\Psi^2\right)S\leq 0$ impose some severe restrictions on the particle creation rate, which must be taken into account when investigating the late evolution of the Universe. In particular, the condition $\ddot{S}=0$ gives $\Psi\propto 1/t$, that is, to a time decreasing particle creation rate, which becomes zero in the very large cosmological time limit.

An interesting and important repercussion of particle creation is its implication on the problem of the arrow of time. The problem of the time arrow consists in finding a mechanism that generates a linear evolution of
time, thus allowing us to differentiate the past of the Universe from its future. There are two different arrows of time. The first is the arrow of time generated thermodynamically and fully determined by the direction in which the entropy of the Universe increases. On the other hand, the cosmological time arrow is determined by the direction in which the Universe expands. Matter production introduces an asymmetry in the Universe’s temporal expansion
and allows us to institute a thermodynamical arrow of time, determined by the particle creation processes. In the cosmological models investigated in the present work, the  arrow
of time defined thermodynamically coincides with the cosmological one, defined by the expansion of the Universe, with both pointing towards an identical arrow of global evolution.

The creation of matter from the cosmological vacuum represents one of the notable results of the quantum theory of fields in curved spacetimes \cite{Q1,Zeld,Q3,Q4,Q5}. The particle production processes may play an important role in the approaches based on quantum field theoretical formalisms to the gravitational interaction, where they emerge naturally. A
crucial conclusion of quantum theory of fields in curved
spacetimes is that in the time-dependent FLRW geometry from the minimally coupled scalar
field, quantum particles are created due to the expansion of the Universe \cite{Q5}. Hence, particle generation processes that naturally appear in quantum theories of gravity, or in quantum field theory in curved spacetimes and in modified gravity theories with torsion--matter coupling, may point towards the possible existence of a profound connection between
these very different approaches for the description of the gravitational interaction. Moreover, one may suggest that modified gravity theories with torsion--matter coupling could provide a compelling phenomenological description of the quantum gravitational processes. As for the nature of the created particles, the present classical macroscopic formalism considered in the present work does not give any insights into the problem. One possibility would be particle creation from geometry in the form of the dark energy particles \cite{BoHa}, having masses of the order of $m=\hbar H_0/c^2\approx 3.8\times 10^{-66}$ g, where $\hbar$ is Planck's constant. Moreover one can assume some similarities between matter creation from geometry and the process of particle creation due to quantum fluctuations, leading to random fluctuations of the physical fields in a small region of the vacuum. Quantum vacuum fluctuations lead to the formation of virtual particles, always created in the form of particle--antiparticle pairs \cite{Q5}. Since virtual particles  are generated spontaneously without a source of energy, vacuum fluctuations and the newly created virtual particles breach the fundamental principle of the conservation of energy. However, this problem may be solvable in geometric theories with geometry--matter coupling, where the source of the (virtual) particles is geometry, and the energy nonconservation has a clear physical origin.

In the present work, we have considered a thermodynamic interpretation of the torsion--matter coupling in the corresponding  class of modified gravity theories, and we have analyzed some of the cosmological implications of this interpretation. The basic tools developed in this approach may be used to further investigate the physical, geometrical, and cosmological properties of   torsion  in the description of gravitational phenomena.

\vspace{6pt}

\authorcontributions{All the authors have substantially contributed to the present work. All authors have read and agreed to the published version of the manuscript.}

\funding{FSNL acknowledges support from the Funda\c{c}\~{a}o para a Ci\^{e}ncia e a Tecnologia (FCT) Scientific Employment Stimulus contract with reference CEECINST/00032/2018, and the research grants No. UID/FIS/04434/2020, No. PTDC/FIS-OUT/29048/2017, and No. CERN/FIS-PAR/0037/2019. }

\institutionalreview{Not applicable.}

\informedconsent{Not applicable.}

\dataavailability{Not applicable. }

\acknowledgments
{We thank the four anonymous reviewers for the careful reading of the manuscript and for comments and suggestions that helped to improve our manuscript.}



\conflictsofinterest{The authors declare no conflict of interest.}

\end{paracol}
\reftitle{References}

\end{document}